\documentclass[twocolumn]{aastex62}

\shortauthors{Berretti et al.}

\begin{document}

\title{Can instrumental effects obscure the true photospheric wave spectrum?}

\correspondingauthor{Michele Berretti}
\email{michele.berretti@unitn.it}

\author[0009-0007-2465-1931]{M. Berretti}
\affiliation{University of Trento \\
Via Calepina 14, 38122 Trento, Italy}
\affiliation{University of Rome Tor Vergata, Department of Physics \\
Via della Ricerca Scientifica 3, 00133 Rome, Italy}

\author{M. Stangalini}
\affiliation{ASI Italian Space Agency \\
Via del Politecnico snc, 00133 Rome, Italy}

\author{D. B. Jess}
\affiliation{Astrophysics Research Centre, School of Mathematics and Physics, Queen’s University Belfast \\
Belfast, BT7 1NN, Northern Ireland, UK}
\affiliation{Department of Physics and Astronomy, California State University Northridge \\
Northridge, CA 91330, USA}

\author{S. Jafarzadeh}
\affiliation{Astrophysics Research Centre, School of Mathematics and Physics, Queen’s University Belfast \\
Belfast, BT7 1NN, Northern Ireland, UK}

\author{S. D. T. Grant}
\affiliation{Astrophysics Research Centre, School of Mathematics and Physics, Queen’s University Belfast \\
Belfast, BT7 1NN, Northern Ireland, UK}

\author{G. Verth}
\affiliation{Plasma Dynamics Group, School of Electrical and Electronic Engineering, The University of Sheffield \\
Mappin Street, Sheffield, S1 3JD, UK}

\author{V. Fedun}
\affiliation{Plasma Dynamics Group, School of Mathematical and Physical Sciences, The University of Sheffield \\
Hicks Building, Hounsfield Road, Sheffield, S3 7RH, UK}

\author{G. Chambers}
\affiliation{Astrophysics Research Centre, School of Mathematics and Physics, Queen’s University Belfast \\
Belfast, BT7 1NN, Northern Ireland, UK}

\author{F. Berrilli}
\affiliation{University of Rome Tor Vergata, Department of Physics \\
Via della Ricerca Scientifica 3, 00133 Rome, Italy}

\begin{abstract}

Optical aberrations and instrument resolution can affect the observed morphological properties of features in the solar atmosphere. However, little work has been done to study the effects of spatial resolution on the dynamical processes occurring in the Sun's atmosphere. In this work, owing to the availability of high-resolution observations of a magnetic pore captured with the Interferometric BIdimensional Spectrometer mounted at the Dunn Solar Telescope, we studied the impact of the diffraction limit and the sampling of an instrument on line-of-sight Doppler velocity oscillations. We reported a noticeable shift in the dominant frequency band from $5$ to $3$ mHz, as both the angular and detector resolutions of the instruments were degraded. We argue that the observed behaviour is a result of the increased contamination of straylight from neighbouring quiet Sun regions, masking the true behaviour of umbral oscillations. These results suggest that the wave energy contributions reported in the literature and based on low-resolution instrumentation may be fundamentally underestimated. As we move into the era of high-resolution instrumentation such as DKIST and MUSE, this work will offer a critical baseline for interpreting new observations, especially in terms of distinguishing true dynamic behaviours from artefacts introduced by instrument-related limitations.

\end{abstract}


\section{Introduction} \label{sec:intro}

When observing the Sun's atmosphere through the lenses of a telescope, we are generally constrained by limitations within both the optical train and the sensor. These can introduce instrumental artefacts, residual optical aberrations, and increased straylight contamination, which may alter the properties of the observed solar plasma and magnetic fields \citep[][]{1992A&A...257L...4D, 1994SoPh..153...91N, 1996ApJ...463..797T, 2007A&A...465..291M, 2011A&A...532A.136S}. Critical factors defining any astronomical observation include the instrument's angular resolution, the point spread function (PSF), the detector sampling (pixel size), and the observing wavelength. For ground-based instruments, atmospheric seeing conditions must also be considered, as they often limit effective resolution and image stability. Furthermore, each instrument is developed with its own unique optical system, making it challenging to compare observations captured from different observatories. 

\begin{figure*}[h!]
    \centering
    \includegraphics{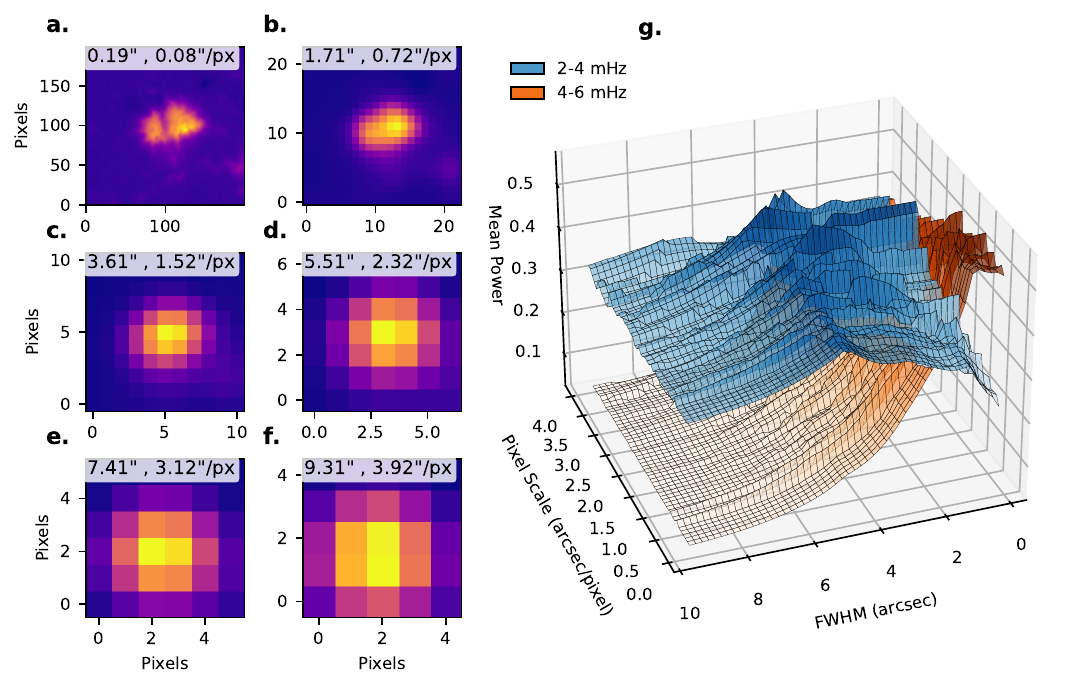}
    \caption{\textbf{Analysis of the power spectral density in the spatial resolution parameters space.} \textbf{a.} Crop of the field of view (FoV) of IBIS centred on the magnetic pore considered in this work at the original resolution. \textbf{b.--f.} Degraded versions of the FoV. The spatial resolution parameters are reported on top of every panel, with the first one referring to the FWHM of the diffraction limited PSF and the second one to the pixel scale. \textbf{g.} 3D surface plots showing the spectral power in the $2-4$~mHz (blue) and $4-6$~mHz (orange) frequency bands as a function of both pixel scale and FWHM. The color gradient on both curves is directly related to the mean power and is only used to improve the visualization.}
    \label{fig:1}
\end{figure*}

Understanding how these effects impact observations of solar plasma and magnetic fields is, therefore, critical for interpreting physical models and reconciling multi-instrument datasets. In \cite{danilovic_intensity_2008}, the authors studied the effects of spatial resolution and straylight on granulation contrast by comparing HINODE/SOT \citep{kosugi_hinode_2007} observations with 3D MHD simulations. In their work, they showed that simulations, once degraded as if they were observed by the telescope, are in nearly perfect agreement with observations and provide a benchmark for the effects of the optical system on the observed solar granulation. Concerning the magnetic field, in \cite{leka_modeling_2012}, the authors degraded synthetic data using four different methods: one method that degrades the Stokes parameters and three methods that directly degrade the images. They compared the results to HINODE/SOT observations. The results showed that the degrading effects of spatial resolution lead to increased magnetic field intensity, lower total flux, and a field vector biassed toward the line of sight. Furthermore, in \cite{milic_spatial_2024}, the authors reported that even spatially averaged quantities, such as open magnetic flux, are affected. 

On the other hand, regarding the effects on the dynamical processes in the solar atmosphere, in \cite{hofmann_effect_1995}, the authors showed for the first time the dependence of the observed amplitude of $p$-modes on the spatial resolution of the instrument by convolving observations captured with the Vacuum Tower Telescope at the Sacramento Peak Observatory with a Gaussian PSF that has a progressively larger FWHM (Full Width at Half-Maximum). In \cite{eklund_sun_2020}, the authors studied the impact of angular resolution on the observability of small-scale dynamics in the solar chromosphere with the Atacama Large Millimetre/sub-millimetre Array (ALMA). Specifically, they used the 3D MHD simulation code Bifrost to produce synthetic observables, which were then degraded to the resolution of each of the configurations of ALMA, studying the impact on their small-scale dynamics. They concluded by demonstrating the crucial role of spatial resolution in observing small-scale dynamics, highlighting the promising capabilities of ALMA for solar observations and how simulations can aid observations by providing correction factors. Furthermore, in \cite{bello_gonzalez_acoustic_2009} and \cite{bello_gonzalez_acoustic_2010}, the authors highlight the importance of spatial resolution for determining the energy flux of acoustic waves. They showed that the total acoustic energy flux was more than twice as large in the latter study, which benefited from higher spatial resolution observations, despite analysing the same atmospheric heights and frequency range. Finally, \cite{jess_reply_2021} noted that for ground-based observatories, seeing conditions can significantly hinder spectral energy at higher frequencies. Given this observational challenge, \cite{jess_waves_2023} provided the first dedicated analysis of the effects of spatial resolution on oscillations detected within magnetic structures, establishing a correlation between the instrument's angular resolution and the measured spectral power at $3$ and $5$~mHz using SST/CRISP observations.
    
In this work, for the first time, we timely bring to the attention of the community the combined impact of both angular and detector resolutions, considering the famous case study of the magnetic pore from 2008, observed with the Interferometric BIdimensional Spectrometer \citep[IBIS,][]{cavallini_ibis_2006}, which showed no power at $3$~mHz and a dominant frequency of 5 mHz. We progressively degraded the angular and detector resolutions of the instrument and studied their effects on the power spectral density of line-of-sight velocity oscillations at two notable frequency bands: the $p$-mode and the $5$~mHz bands.
    
\section{Dataset} \label{sec:dataset}

The data used in this work were captured with IBIS mounted at the Dunn Solar Telescope (New Mexico, USA) on 2008 October 15 at $16:30$~UTC. IBIS is a multi Fabry-Perot interferometer designed to capture high spatial resolution monochromatic images of the Sun's atmosphere. The target of this observation was a small magnetic pore with a light bridge slightly off-centre at [25.2 N, 10.0 W] on the solar disk. 

The dataset includes a sequence of $80$ full Stokes spectral scans across the Fe~{\sc i}~$617.3$~nm photospheric absorption line, with a temporal cadence of $52$ seconds between successive scans. The spatial sampling of the detector is $0{\,}.{\!\!}{''}08$ per pixel, while the angular resolution of the instrument is $0{\,}.{\!\!}{''}16$. Furthermore, it is equipped with an Adaptive Optics system to reduce the effects of seeing conditions. The obtained spectropolarimetric images were then subjected to a rigorous calibration process, for which we refer the reader to \cite{reardon_characterization_2008}, which included the deconvolution of the IBIS PSF. Post-calibration, images were co-registered and de-stretched to minimise residual seeing effects.

Line-of-sight Doppler velocity was estimated following the procedure highlighted in \cite{schlichenmaier_flow_2000}, essentially tracking the minimum of the line profile as in a Fourier tachometer. The line-of-sight magnetic field and its inclination angle ($\gamma$) maps were obtained using the SPINOR inversion code \citep{frutiger_properties_2000}. 

\section{Methods \& Results}

Both the Doppler velocity and line-of-sight magnetic field maps were degraded to emulate what a telescope with progressively lower angular and detector resolutions would observe. To degrade the angular resolution of the instrument, we defined a synthetic PSF based on an Airy disc with a varying FWHM depending on the resolution we wished to achieve. To simulate lower detector resolutions, we binned the original pixels into square blocks. The value of each new, larger pixel was estimated considering the intensity-weighted average of the values of the pixels it comprises. As a pixel gets larger, it inherently collects more light, increasing the signal-to-noise ratio, but the resulting magnetic field and doppler velocity values represent the mean physical properties of the area, weighted by the intensity of the contributing structures. In total, we degraded the angular and detector resolutions, starting from the original IBIS values down to $5{\,}.{\!\!}{''}7$ per pixel and $2{\,}.{\!\!}{''}4$ per pixel, respectively. This process resulted in a two-dimensional parameter space representing various combinations of the two spatial resolution parameters. In Fig. \ref{fig:1} panels a. through f., the initial frame of the $B_{los}$ data cube is presented for six different points along the bisector of this two-dimensional parameter plane.

For each pair of angular and detector resolution, we selected regions where $B_{los}$ is above 75\% of the $99^{th}$ percentile. Within these regions, we computed the periodogram of the line-of-sight velocity time series for each pixel and averaged it to create a single mean power spectrum per region. To estimate the power spectra of the LOS velocity time series, we followed the procedure highlighted in \cite{chatfield_analysis_2003}, \cite{jess_waves_2023}, and  \cite{jafarzadeh_wave_2025}. We then normalised each mean spectrum and linked the spatial resolution parameters to the respective averaged power spectral densities in the $2.5-3.5$~mHz and $4.5-5.5$~mHz frequency bands. 

In Fig.~\ref{fig:1} panel g., we show the impact of degraded spatial resolution on the two most frequently considered frequency bands when dealing with waves and wave propagation in the solar atmosphere: the photospheric $p$-modes/magnetoacoustic waves, centred at $3$ mHz, and the usually chromospheric one centred at $5$~mHz. Specifically, we show the power spectral density of line-of-sight velocity oscillations across the 2D resolution parameter space for the $2.5-3.5$~mHz and $4.5-5.5$~mHz bands. At the highest resolutions, power is concentrated in the 5 mHz band. As the resolution decreases, however, this power transfers significantly to the 3 mHz band. In Appendix~\ref{ap:2}, we show the power spectra sensitivity to a varying pixel scale while the FWHM is held constant at two different values. Conversely, in Appendix~\ref{ap:3} we show the sensitivity to a varying FWHM while the pixel scale is fixed at two values.

\begin{figure}[h!]
    \centering
    \includegraphics{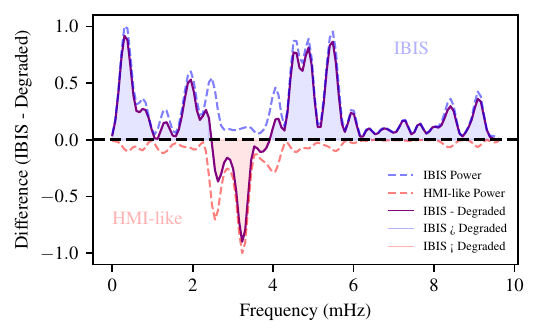}
    \caption{A comparison between the IBIS power spectrum (blue, dashed) and the HMI-like degraded one (orange, dashed; inverted for clarity). The solid violet line represents their difference, while the shaded blue and red regions highlight the frequency domains of higher sensitivity for IBIS and HMI, respectively.}
    \label{fig:2}
\end{figure}

To place our results in the context of other existing and well-established instruments, we degraded our IBIS dataset to simulate observations from HMI. This process involved convolving our data with the appropriate PSF model, estimated in \cite{yeo_point_2014} thanks to the transit of Venus, and resampling it to the pixel size of the target instrument, which is equal to $0{\,}.{\!\!}{''}5$ per pixel. In Fig.~\ref{fig:2}, we show the difference between the original IBIS power spectrum and the HMI-like degraded one. This difference shows which frequencies dominate the observed power spectra at a given set of resolution parameters. Specifically, the power in the original IBIS spectrum is mostly confined to the high-frequency regime, with dominant peaks between $5$ and $6$~mHz. On the other hand, the HMI-like power spectrum is dominated by power at $3$~mHz.

\begin{figure}[h!]
    \centering
    \includegraphics{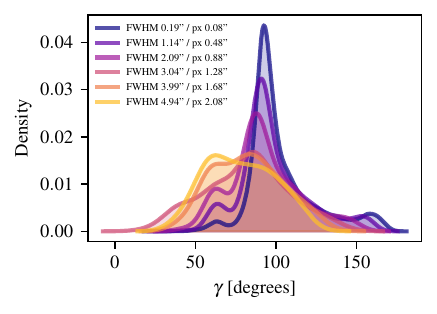}
    \caption{Distributions of the inclination angle $\gamma$ for different degradation steps on the bisector of the spatial resolution parameters space, from violet to yellow as shown in the legend.}
    \label{fig:3}
\end{figure}

Finally, the effect of degraded resolution is evident in the magnetic field inclination, $\gamma$. As shown in Fig.~\ref{fig:3}, the distribution of $\gamma$ becomes progressively more uniform as the resolution decreases. This loss of structural detail implies that the magnetic features of the umbra are lost, blending their properties with those of the neighbouring quiet Sun. While this blending can be easily foreseen as we degrade the spatial resolution, we have quantitatively shown here how easily instrumental limitations can bias the retrieved magnetic field topology, serving as a warning against over-interpreting magnetic field inclination values obtained from inversions of unresolved structures.
    
\section{Discussions}

The generally accepted view when discussing waves in the solar atmosphere, and specifically their frequencies, is that $p$-modes dominate ubiquitously in the photosphere at $3$~mHz \citep{khomenko_oscillations_2015, jess_waves_2023}, with the same frequency then being injected into the magnetic structures by a mechanism known as mode conversion \citep{spruit_conversion_1992, dsilva_acoustic_1994}. As we move up, due to the action of an acoustic cut-off, this frequency shifts from $3$ to $5$~mHz \citep{felipe_multi-layer_2010, felipe_height_2018}. However, in the last decade, a series of works have started to challenge this interpretation.

In \cite{stangalini_three-minute_2012}, the authors reported the peculiar case of a pore observed with IBIS at the DST, showing, at a photospheric height, dominant power at $5$~mHz and little to no spectral power at $3$~mHz. However, the authors lacked the statistical consistency required to discern whether the observed behaviour was restricted to this case or represented a more general pattern. In \cite{stangalini_novel_2021}, the authors, using the same pore dataset again, showed that the regions with the highest amount of spectral power at $5$~mHz were constrained within the magnetic structure, leaving the non-magnetised regions of the FoV at $3$~mHz, as expected. In an effort to provide statistical consistency, in \cite{berretti_unexpected_2024}, the authors studied the horizontal velocity oscillations of more than one million small-scale magnetic structures in the photosphere. Despite the different nature of the studied oscillations, they found a shared dominant frequency of $5$~mHz across all the differently shaped and sized elements. This is a clear first indication of a common driver that excites these oscillations, which led the same authors to study the line-of-sight velocity oscillations of nearly one thousand sunspots across one full solar cycle in \cite{berretti_umbral_2025}, where they found a closely packed series of peaks within $4$ and $6$~mHz. 

Here, in an attempt to shed further light on the latest findings, we studied the extent to which spatial resolution plays a role in the ability to detect higher frequency oscillations and introduces observational artefacts. In Fig.~\ref{fig:2}, we have shown that the detected spectral power strongly depends on both the angular and detector resolutions of the instrument, as a significant decrease in the power spectral density in the $5-6$~mHz frequency band is measured when the resolutions are progressively degraded. Indeed, the broadening of the PSF causes the signal from the highly magnetised and thus photon-starved regions of the FoV to become increasingly contaminated by straylight from the nearby quiet Sun. This interpretation is strengthened by the magnetic field inclination distributions, which reveal a lower contrast between the umbra of the pore and the nearby quiet Sun, leading to a unified behaviour across the whole FoV. Furthermore, insufficient resolution prevents the detection of the low-amplitude motions that result from the line-of-sight velocity oscillations. Indeed, if the displacement of the longitudinal oscillation is smaller than the pixel scale, the motion becomes spatially unresolved. 

Moreover, we argue that the unphysical ridge observed at $0{\,}.{\!\!}{''}16$ per pixel is not a physical structure but an artefact of spatial resonance, occurring as the degraded pixel scale matches the instrument's true angular resolution. An additional resonance effect is expected when the pixel size approaches the diameter of the magnetic pore, causing another enhancement in spectral power, followed by a sudden drop as the pixels become larger. This effect is visible in Fig.~\ref{fig:a1} of Appendix~\ref{ap:1}.

Despite the slightly different frame rates of the two instruments, the impact of the degraded resolution is clear from Fig.~\ref{fig:2}. Apart from a much smaller peak at $4$~mHz, none of the higher frequency features observed in the IBIS spectrum survived in the HMI-like spectrum. Surprisingly, the peak at $3$ mHz, which is the most prominent feature of the degraded spectrum, is non-existent in the original data, suggesting that the $3$~mHz oscillation is measured in regions of the FoV outside the umbra of the pore, thus strengthening our interpretation of the effects of spatial resolution and enhanced straylight. Finally, the HMI-like power spectrum is similar to observations of more than $600$ sunspots with the actual HMI instrument in \cite{berretti_umbral_2025}, providing a reassuring further check on our analysis.

\section{Conclusions}

In this work, we have studied the effects that the angular resolution and pixel size of an instrument may have on the observed line-of-sight velocity oscillations within a magnetic structure. As a case-study, we considered the magnetic pore observed with IBIS in 2007, which has already been featured in a plethora of other works and is known for showing no power at $3$ mHz at photospheric heights, but rather a dominant frequency of $5$~mHz more commonly associated with chromospheric heights. We have shown a clear correlation between the observed power in the two most commonly considered frequency bands (i.e., the $3$~mHz and $5$~mHz centred ones) and the spatial resolution of the instrument, demonstrating how, at lower resolutions, higher frequencies are effectively damped due to both the significantly higher susceptibility to straylight and the instrument inability to resolve the amplitude of such oscillations. Our work stresses the importance of accounting for the basic optical properties of the instrument when studying the dynamical properties of magnetic structures in the Sun's atmosphere and their energy budget. Indeed, estimations of the wave energy contribution in magnetic structures, which are often based on low-resolution data, might be underestimated and could require significant reconsideration. Indeed, while it is well-established that magnetic structures host both $3$ and $5$~mHz oscillations, the relative power between the two has never been robustly established. Here, we wish to highlight the importance of understanding the instrumental artefacts introduced by the spatial resolution, which necessitates a careful assessment of instrument capabilities when analysing observed oscillations. The immediate next step should involve a comprehensive programme of cross-observations utilising currently available solar observatories.

\acknowledgments

 MB acknowledges that this publication (communication/thesis/article, etc.) was produced while attending the PhD program in PhD in Space Science and Technology at the University of Trento, Cycle XXXIX, with the support of a scholarship financed by the Ministerial Decree no. 118 of 2nd March 2023, based on the NRRP - funded by the European Union - NextGenerationEU - Mission 4 "Education and Research", Component 1 "Enhancement of the offer of educational services: from nurseries to universities” - Investment 4.1 “Extension of the number of research doctorates and innovative doctorates for public administration and cultural heritage” - CUP E66E23000110001. We wish to acknowledge scientific discussions with the Waves in the Lower Solar Atmosphere (WaLSA; \href{https://WaLSA.team}{www.WaLSA.team}) team, which has been supported by the Research Council of Norway (project no. 262622), The Royal Society (award no. Hooke18b/SCTM; \citealt{2021RSPTA.37900169J}), and the International Space Science Institute (ISSI Team 502). DBJ acknowledge support from the Leverhulme Trust via the Research Project Grant RPG-2019-371. DBJ and SJ wish to thank the UK Science and Technology Facilities Council (STFC) for the consolidated grants ST/T00021X/1 and ST/X000923/1. DBJ and SDTG also acknowledge funding from the UK Space Agency via the National Space Technology Programme (grant SSc-009). SJ also acknowledges support from the Rosseland Centre for Solar Physics, University of Oslo, Norway, and the Max Planck Institute for Solar System Research, Germany. VF and GV are grateful to the Science and Technology Facilities Council (STFC) grants ST/V000977/1, ST/Y001532/1, and The Royal Society, IEC/R3/233017 - International Exchanges 2023 Cost Share (NSTC), collaboration with Taiwan. VF would like to thank the International Space Science Institute (ISSI) in Bern, Switzerland, for the hospitality provided to the members of the team on ``Opening new avenues in identifying coherent structures and transport barriers in the magnetised solar plasma''.

\vspace{5mm}
\facilities{DST (IBIS)}

\bibliography{references, references_zotero}
\bibliographystyle{aasjournal}

\clearpage

\appendix

\section{Sensitivity to higher pixel scales}\label{ap:1}

\begin{figure*}[h!]
    \centering
    \includegraphics{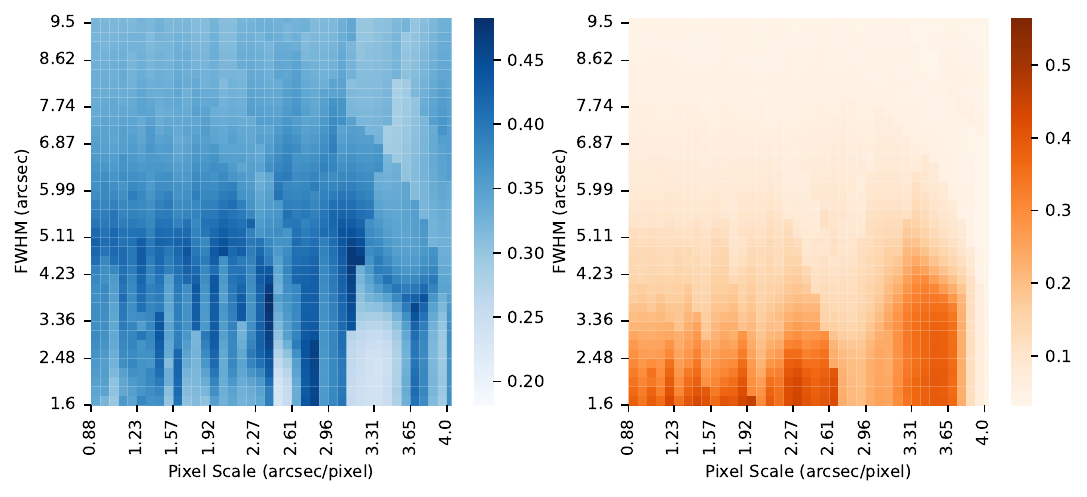}
    \caption{\textbf{Two-dimensional mapping of spectral power sensitivity.} Heatmaps of the spectral power as a function of both angular resolution and pixel scale at $3$~mHz for the left panel and $5$~mHz for the right one. The structure shows the expected power enhancement and the following drop when the pixel scale matches the characteristic physical length scale of the pore.}
    \label{fig:a1}
\end{figure*}

In our analysis, two self-sustained checks can be performed to assess the robustness of our work. One, presented in the main body, is the spatial resonance observed when the degraded pixel scale matches the true angular resolution of the instrument. A second check is instead at much higher pixel sizes. Indeed, when the pixel is just about as big as the umbra of the pore, we should see an enhancement at the frequencies of the original power spectrum, followed by a sharp drop as the pixels become larger than the magnetic structure itself. In Fig.~\ref{fig:a1}, we show the results of this second check.

\clearpage

\section{Sensitivity to pixel scale at fixed angular resolutions}\label{ap:2}

\begin{figure}[ht!]
    \centering
    \includegraphics{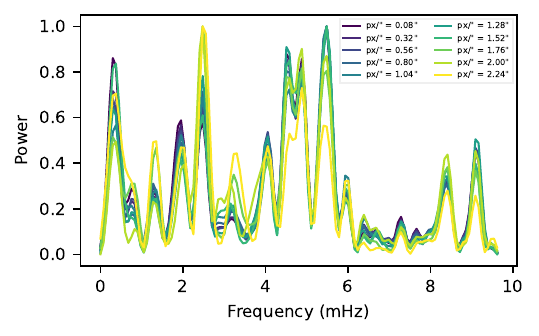}
    \includegraphics{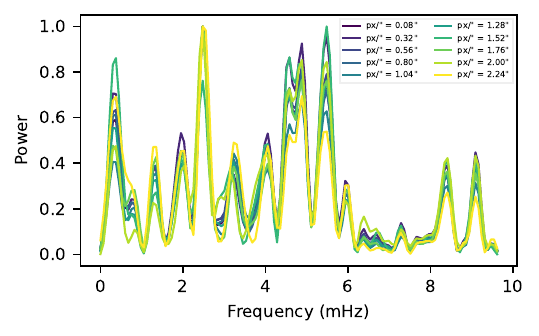}
    \caption{\textbf{Effect of varying pixel scale on power spectra for two distinct fixed angular resolutions.} Power spectra as a function of frequency for several pixel scales (as shown in the legend) at a constant angular resolution. The upper panel is fixed at FWHM = $0{\,}.{\!\!}{''}19$, and the lower panel is fixed at FWHM = $1{\,}.{\!\!}{''}52$.}
    \label{fig:b1}
\end{figure}

In the left and right panels of Fig.~\ref{fig:b1}, we show how the power spectra change as we degrade the pixel scale while keeping the angular resolution fixed at $0{\,}.{\!\!}{''}19$ and $1{\,}.{\!\!}{''}52$, respectively.

\clearpage

\section{Sensitivity to angular resolution at fixed pixel scales}\label{ap:3}

\begin{figure}[ht!]
    \centering
    \includegraphics{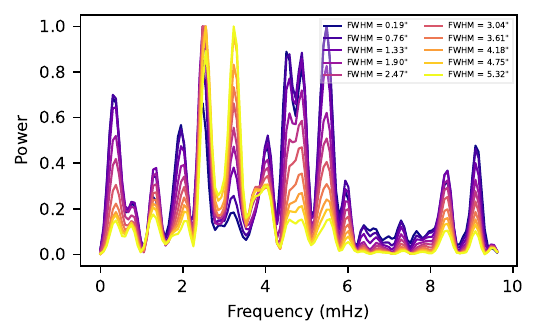}
    \includegraphics{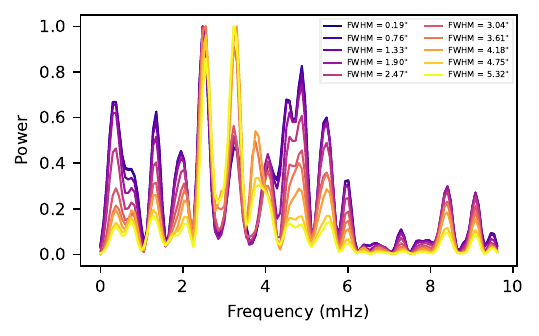}
    \caption{\textbf{Effect of varying angular resolution on power spectra for two distinct fixed pixel scales.} Power spectra as a function of frequency for several angular resolutions (as shown in the legend) at a constant pixel scale. The upper panel is fixed at $0.24px/"$, and the lower panel is fixed at $1.84px/"$.}
    \label{fig:c1}
\end{figure}

In the left and right panels of Fig.~\ref{fig:c1}, we show how the power spectra change as we degrade the angular resolution, keeping the pixel scale fixed at $0.24px/"$ and $1.84px/"$, respectively.

\end{document}